\documentclass[epj]{svjour}
\usepackage{latexsym}
\usepackage{graphics}
\usepackage{graphicx}
\usepackage{epsfig}
\usepackage{epsf}

\def\slash#1{\setbox0=\hbox{$#1$}
   \dimen0=\wd0 \setbox1=\hbox{/} \dimen1=\wd1
   \ifdim\dimen0>\dimen1 \rlap{\hbox to \dimen0{\hfil/\hfil}} #1
   \else  \rlap{\hbox to \dimen1{\hfil$#1$\hfil}} / \fi}
\def\intk{\int \frac{d^4 k}{(2\pi)^4}}
\def\D{{\bf D}}

\begin{document}
\title{Relativistic NJL Model with Light and Heavy Quarks }
\author{\underline{Andr\'{e} Luiz Mota} \inst{1} \inst{2}
\thanks{\emph{Speaker at QNP06 Madrid , 5-10 June 2006. Work supported
by CAPES-Brazil and Spanish DGI and FEDER funds with grant
no. BFM2002-03218, Junta de Andaluc\'{\i}a grant No. FQM-225, and EU
RTN Contract CT2002-0311 (EURIDICE) } } \and Enrique Ruiz Arriola
\inst{2}
}                     
\institute{
Departamento de Ci\^{e}ncias Naturais,
Universidade Federal de S\~{a}o Jo\~{a}o del Rei,
36301-160, S\~{a}o Jo\~{a}o del Rei,
 Brazil
\and 
Departamento de F\'{\i}sica At\'omica, Molecular y
Nuclear, Universidad de Granada, E-18071 Granada, Spain
}
\date{\today}
%
\abstract{We study the Nambu-Jona-Lasinio model with light and heavy
quarks in a relativistic approach. We emphasize relevant
regularization issues as well as the transition from light to heavy
quarks. The approach of the electromagnetic meson form factor to the
Isgur-Wise function in the heavy quark limit is also discussed.
} 
\maketitle

The physics of light and heavy quarks and their corresponding
effective field theories cannot be more disparate even though a smooth
transition between both limits is expected. In the case of light up,
down and strange quarks the spontaneous breaking of chiral symmetry is
the dominant feature which explains the mass gap between pions and
kaons and the rest of the hadronic spectrum enabling the use of Chiral
Perturbation Theory (ChPT) for energies much smaller than the mass
gap~\cite{Gasser:1983yg,Donoghue:1992dd}. In the opposite limit of heavy charm,
bottom and top quarks, spin symmetry largely explains the degeneracy
between hadronic states which differ only in the spin of the heavy
quark like e.g. $B(5280) $ vs $B^*(5325) $, or $D(1870)$ vs
$D^*(2010)$ and a systematic Heavy Quark Effective Theory
(HQET)~\cite{Georgi:1990um,Neubert:1993mb,Manohar:2000dt} can be
designed for masses much larger than the mass gap. Besides these two
fairly known extreme limits, the understanding of the transition from
light to heavy quarks is not only of theoretical interest, but may
also provide some insight into lattice simulations where the putative
light quarks are most frequently artificially heavy. Unfortunately,
there is no general framework describing the heavy-light transition in
a model independent way, even though ChPT and HQET describe the
extreme cases.

In HQET the heavy quark limit is taken {\it before} implementing
dimensional regularization because as is well known heavy particles do
not decouple in this regularization scheme. For a heavy quark the
relevant degrees of freedom are given by
\begin{equation}
\Psi_v(x)=\frac{1+\not{v}}{2} e^{i m_0 v.x} \Psi(x)
\end{equation}
where $\Psi(x)$ is the heavy quark spinor, $m_0$ is the heavy quark
mass, and $v^{\mu}$ is a quadrivector where the spacial components
$\vec{v}$ corresponds to the velocity of the heavy quark and the time
component is chosen in order to have $v^2=1$. After integrating out the irrelevant
degrees of freedom, the resulting effective
Lagrangian is expanded in $1/m_0$, and the propagator for the heavy
quark effective field, in leading order, is given by
\begin{equation}
S(k)=\frac{1}{v.k+i \epsilon},
\end{equation}
$k^{\mu}$ being the residual momentum of a heavy quark with total
momentum $k^{\mu} + m_0 v^{\mu}$.

In this work we discuss the heavy-light transition with the guidance
of the Nambu--Jona-Lasinio (NJL) model for quarks (for reviews see
e.g. \cite{Vogl:1991qt,Klevansky:1992qe,Hatsuda:1994pi,Christov:1995vm}).
The corresponding Lagrangian reads
\begin{eqnarray}
{\cal L}=\bar{\psi} (i \slash{\partial} - \hat{m}_0) \psi  
-\frac{G}{2}( (\bar{\psi} \lambda \psi)^2 + (\bar{\psi} i \gamma_5 \lambda \psi)^2 ) 
\label{Snjl}
\end{eqnarray}
where $\lambda$ are the $N_f^2-1$ flavour $SU(N_f)$ Gell-Mann matrices
and $\hat{m}_0={\rm diag}(m_{u0},m_{d0},m_{s0},...,m_{n0})$ is a
diagonal current mass matrix which explicitly breaks chiral
invariance. With the exception of the mass term all flavours are
treated on the {\it same footing} and Lagrangian~(\ref{Snjl}) is
invariant under the $SU_R (N_f) \otimes SU_L(N_f) $ chiral group and
also under $SU(N_c)$ global transformations. Summation over color and
flavour indexes is implicit.  As it is well known the NJL model is not
renormalizable and a finite cut-off $\Lambda$ is required to make
sense of it. A technical, but crucial, issue is that of the finite
cut-off regularization method and its consistency with the gauge and
chiral symmetries.

Quark models have been studied in the last decade in connection to the
heavy quark effective Lagrangian obtained from HQET and its interplay
with chiral quark
models~\cite{Ebert:1994tv,Deandrea:1998uz,Hiorth:2002pp,Bardeen:2003kt}.
The basic underlying assumption is that all quark species are assumed
to have the same kind of contact interactions and the mass terms are
indeed treated in a rather asymmetric fashion. However, mimicking HQET
itself, in HQET models, the infinite heavy quark limit is taken {\it
before} applying a definite regularization scheme in the heavy
sector. Whether or not the $m_0 \rightarrow \infty$ limit commutes
with the regularization procedure is not obvious. In addition, due to
the non-renormalizability of the models, as in the light quarks NJL
model, the finite regularization procedure employed is a part of the
model and different regularizations can lead to different results. The
choice of a regularization procedure that does not violate symmetries
of the model becomes relevant, and in particular naively shifting the
internal momenta of the loops may be dangerous. There is no reason
{\it a priori} why neglecting finite cut-off corrections is more
justified in the heavy sector as it is in the light sector.

Following previous experience we use the Pauli-Villars method with two
subtractions in the coincidence limit. The proper way to do this is by
bosonization and separation of the effective action into normal and
abnormal parity contributions~\cite{Schuren:1991sc}. After integrating
out the fermions one gets the normal parity contribution to the
effective action
\begin{eqnarray}
S_{\rm even} &=& -\frac{i N_c}{2} \sum_i c_i {\rm Tr} \log (\D_5 \D+
\Lambda_i^2) \nonumber \\ &-& \frac{1}{4 G_S}\int d^4 x {\rm tr}_f
(S^2+P^2) ,
\label{efac}
\end{eqnarray}
where $\D$ and $\D_5$ are  Dirac operators given by
\begin{eqnarray}
\D &=& 
+ i \slash{\partial} - S - \hat{m}_0 - i \gamma_5 P + \slash{v} + \slash{a} \gamma_5 \nonumber \\ 
\D_5 &=& -i \slash{\partial} - S - \hat{m}_0+ i \gamma_5 P -
\slash{v} + \slash{a} \gamma_5
\end{eqnarray}
where $S=\lambda^a S^a$, $P=\lambda^a P^a$ are dynamical fields and
$v^{\mu}=\lambda^a v^{\mu}_a$ and $a^{\mu}=\lambda^a a^{\mu}_a$ stand
for external sources. The Pauli-Villars regulators fulfill $c_0=1$,
$\Lambda_0 =0$ and the conditions $\sum_i c_i =0$, $\sum_i c_i
\Lambda_i^2 =0 $ which render finite the logarithmic and quadratic
divergences respectively. In practice, we take two cut-offs in the
coincidence limit $ \Lambda_1 \to \Lambda_2 = \Lambda $ and hence $
\sum_i c_i f(\Lambda_i^2)=f(0)-f(\Lambda^2)+\Lambda^2
f^\prime(\Lambda^2) $. For light quarks with small current masses,
$m_0$, the dynamical breaking of chiral symmetry generates a
constituent mass, $M$, for the quarks, and pion physics phenomenology
yields values $\Lambda \sim 1{\rm GeV}$ for $M \sim 300 {\rm MeV}$ and
both the constituent as well as the current masses are much smaller
than the cut-off $m_0 \ll M \ll \Lambda$.

Naively, one would expect that, as a matter of principle, processes
involving scales above the cut-off cannot be reliably addressed by the
model. However, this is not necessarily so.  A compelling example is
provided by the study of high energy processes which involve
asymptotically large momenta $Q^2 \gg \Lambda^2 $ which enable the
determination of mesonic parton leading twist distributions and
amplitudes~\cite{RuizArriola:2002wr}. The surprisingly good agreement
found in such an analysis, at least for the pseudoscalar bosons, when
gluonic radiative corrections via QCD evolution equations are
implemented, suggests not only that there is nothing fundamentally
wrong in looking at high scales as compared to the model cut-off but
also that a rather acceptable description of existing data may be
achieved. An important lesson learned from these studies was that a
sloppy treatment of the finite cut-off regularization violates
significantly relevant constraints regarding gauge and relativistic
invariances which control the normalization and momentum fraction
shared by the constituents respectively. With this insights in mind we
dare to explore with the necessary provisos the NJL model for {\it any
current quark masses} including as a particular case the heavy quark
limit, i.e. for current quark masses much larger than the cut-off $m_0
\gg \Lambda$.

The relevant observable quantities can be read off from the effective
action, by collecting the coefficients of the corresponding terms. The
electroweak decay constant appears as the coefficient of the term
involving one axial vector current and one pseudoscalar meson
field. From now on, we will use $m$ to denote the total 
mass of a light quark and $m_0$ for the total mass of a heavy quark.
For a given channel involving one light quark  and
one heavy quark, one has (PV regularization over-understood)
\begin{eqnarray}
p^{\mu}f_{M}(p^2)&=&i \intk {\rm Tr} \Big\{ i \gamma_{5}
 \frac{i}{(\slash{k} + m_0 \slash{v}) - m_0} \times \nonumber \\ & &
 \qquad \gamma^{\mu} \gamma_{5} \frac{i}{(\slash{k}-\slash{p})-m} \Big\},
\end{eqnarray}
and $f_{M}(M_{\Phi}^2)$ is the heavy meson electroweak decay
constant, where $M_{\Phi}$ stands for the heavy-light meson mass.

In the heavy quark limit, the Isgur-Wise function is a universal form
factor, defined as the matrix elements of the electroweak
heavy-to-heavy currents between two heavy mesons of different
non-relativistic velocities~\cite{IsgurWise}.  
Within the Nambu-Jona-Lasinio model with
heavy quarks, it can be computed as the heavy quark limit of the
electromagnetic form factor for an arbitrary current quark mass,
yielding
\begin{equation}
\Gamma^{\mu}(p,p')=-i(\Gamma_0^{\mu}(p,p')+\Gamma_0^{\mu}(-p',-p)), \label{ffactor}
\end{equation}
where, formally
\begin{eqnarray}
&&\Gamma_0^{\mu}(p,p')=\intk Tr\Big\{ \frac{\slash{k}-m}{k^2-m^2}
\times \label{formfactor} \\ &&\frac{ \slash{k}+m_0
\slash{v}+\slash{p}+m_0}{(k+m_0 v+p)^2 - m_0^2}\gamma^{\mu}
\frac{\slash{k}+m_0 \slash{v}+\slash{p}'+m_0}{(k+m_0 v+p')^2-m_0^2}\Big\}
. \nonumber
\end{eqnarray}
The form factor is computed with on-shell mesons ($p^2=p'^2=(\Delta
M)^2$), $\Delta M = M_{\Phi} - m_0$ and we choose the arbitrary
parameter $v$ such as $v=\frac{p}{\sqrt{p^2}}$. We also define
$\omega=\frac{p.p'}{(\Delta M)^2}$.  As before, we are assuming a
Pauli-Villars gauge invariant regularization scheme. The explicit
result will be given elsewhere~\cite{inpreparation}.

\begin{figure}[ttt]
\vspace{0.3mm}
\begin{center}
\includegraphics[angle=0,width=0.48\textwidth]{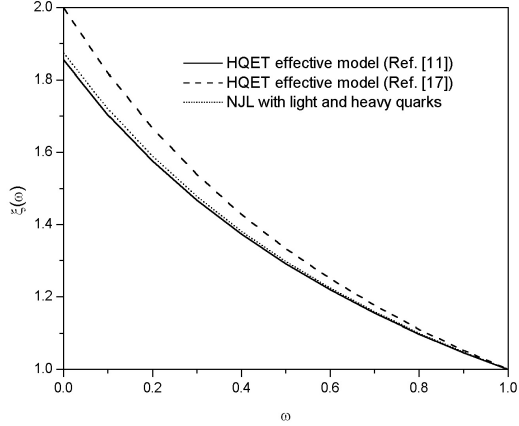}
\end{center}
\vspace{0.3mm}
\caption{ The Isgur-Wise function for the NJL model and other HQET
effective models (Solid line - Ref. \cite{Ebert:1994tv}, dashed line - 
Ref. \cite{Bardeen:1993ae} (with $m=0$), dotted line - present approach.)}
\label{figiwf1}
\end{figure}


\begin{figure}[tb]
\vspace{0.3mm}
\begin{center}
\includegraphics[angle=0,width=0.48\textwidth]{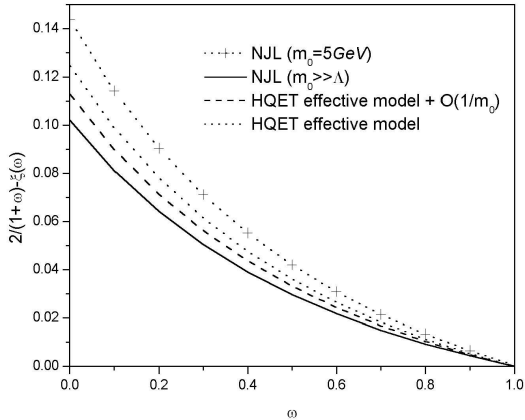}
\end{center}
\vspace{0.3mm}
\caption{Effects of $\frac{1}{m_0}$ corrections to the Isgur-Wise
function. The present results are represented by the solid ($m_0
\rightarrow \infty$) and dotted with crosses ($m_0=5 {\rm GeV}$) lines. The
HQET effective model \cite{Ebert:1994tv} results are represented by
the dotted ($m_0 \rightarrow \infty$) and dashed ($m_0=5 {\rm GeV}$) lines.}
\label{figiwf3}
\end{figure}

In Fig.~\ref{figiwf1} we compare our result to that of Ebert et
al.~\cite{Ebert:1994tv}.  To see clearly the effect when comparing to
, we choose the set of parameters to reproduce $f_{\pi }=93{\rm MeV}$,
$m_{\pi }=140{\rm MeV}$, $m_{\rho }=770{\rm MeV}$ and $M_{B}=5.3{\rm
GeV}$, obtaining $ m_{u}=m_{d}=300{\rm MeV}$, $\Lambda =875{\rm MeV}$,
$m_{s}=510{\rm MeV}$, $m_{B}=5.1{\rm GeV}$ and $G=2.9{\rm GeV^{-2}}$.
In contrast with \cite{Ebert:1994tv}, no independent coupling for the
heavy meson sector needs to be inserted on the model in order to
reproduce the light-light and heavy-light mesons masses, although a
very low $B$ meson weak decay constant is obtained ($f_B=59{\rm
MeV}$). Nevertheless, other processes that could be important to the
correct description of this decay constant, as mesons loops, are
absent in the present treatment. In the same figure, we also compare
to the HQET effective model presented in \cite{Bardeen:1993ae} where
\begin{equation}
\xi (\omega)=\frac{2}{1+\omega }. \label{xi}
\end{equation}
in both $m_0 \rightarrow \infty$ and $m \rightarrow 0$ limits.  As we
see, differences become more significant as $\omega$ goes to zero. The
results obtained from the present model are very close to the results
obtained in \cite{Ebert:1994tv}. The differences of both models to the
results presented on \cite{Bardeen:1993ae} are due to finite light
{\it constituent} quark mass effects: even in the $m_0 \rightarrow \infty$
limit, result (\ref{xi}) is only achieved when $m = 0$. 

In Fig. \ref{figiwf3} we present the comparison between the
behaviour of the Isgur-Wise function as $m_0$ goes from $5 {\rm GeV}$ to $50
{\rm GeV}$ ($m_0 \gg \Lambda$), computed on both present and
\cite{Ebert:1994tv} models (corrections of order $1/m_0$ on the HQET
effective model of Ref. \cite{Ebert:1994tv} were included). As can be
viewed from Fig. \ref{figiwf3}, as $m_0$ increases the slope of the
Isgur-Wise function on the present model increases, while the same
slope decreases on the model presented on \cite{Ebert:1994tv}.  The
magnitude of the changes on the slope of the Isgur-Wise function is
also different in the two models. 

Finally, let us mention that a derivative expansion of the bosonized
version of the model can be employed to construct an effective mesonic
Lagrangian, as was done for the light quark sector
\cite{Ebert:1985kz}. One problem is that such an expansion assumes
small momenta for the corresponding meson bosonized fields, while for
heavy mesons they are large.  It is possible, however, to overcome
this problem in a way that the treatment of light and heavy mesons is
as symmetric as possible, so that at any stage the heavy pseudoscalars
would become Goldstone bosons if the quarks were light. Further
details will be further elaborated elsewhere~\cite{inpreparation}.

\end{document}